\documentclass[twocolumn]{aastex701}

	\usepackage{amsmath}
	\usepackage{amssymb}
	\usepackage{bm}
    \usepackage{listings}
    \usepackage{enumitem}
	
	\let\vec\bm

	\newcommand{\diff}{\ensuremath{\mathrm{d}}}
	\newcommand{\e}{\mathrm{e}}
	\newcommand{\I}{\mathrm{i}}

	\newcommand{\p}{\prime}

	\newcommand{\erfc}{\mathrm{erfc}}

	\newcommand{\dc}{\delta_\mathrm{c}}

	\newcommand{\Msol}{\mathrm{M}_\odot}

	\graphicspath{{./}{figures/}}
		
\begin{document}
			
			\title{The pre-infall bias of subhalos}

			\author[orcid=0000-0003-3808-5321,sname='Delos']{M. Sten Delos}
			\affiliation{Carnegie Observatories, 813 Santa Barbara Street, Pasadena, CA 91101, USA}
			\email[show]{mdelos@carnegiescience.edu}  
			
			\author[orcid=0000-0001-5501-6008,sname='Benson']{Andrew Benson}
			\affiliation{Carnegie Observatories, 813 Santa Barbara Street, Pasadena, CA 91101, USA}
			\email{abenson@carnegiescience.edu}  
			
			\begin{abstract}
				
				Dark matter halos destined to fall into a more massive host differ from typical field halos of the same mass even before infall. In cosmological simulations, we find that the progenitor mass functions of these ``future subhalos'' are systematically shifted toward higher masses, with the shift growing as infall approaches. The bias takes a compact form within extended Press-Schechter theory: the collapse barrier is multiplied by a function $\beta(D/D_\mathrm{infall},a)$, where $D$ is the linear growth factor at scale factor $a$ and $D_\mathrm{infall}$ is the growth factor at infall. We find $\beta(x,a)=(1-x)^{1.20+0.14a}$ for the $M_{200\mathrm{c}}$ mass definition and $(1-x)^{1.20+0.05a}$ for $M_{200\mathrm{m}}$; the explicit scale-factor dependence captures the late-time influence of dark energy. One consequence is that halos shortly before infall are 10--15\% more centrally concentrated than typical field halos of the same mass.
				
			\end{abstract}
			
		
		
		\section{Introduction}\label{sec:intro}

        Dark matter subhalos -- halos that orbit within larger host halos -- differ from field halos of the same mass \citep{2007ApJ...667..859D,2008MNRAS.391.1685S,2017MNRAS.466.4974M}. After a subhalo falls into a host, tidal forces reshape its structure while stripping its mass, and its growth by accretion is suppressed or halted. These post-infall effects have been extensively studied in both simulations and analytics, and they are a key ingredient in semianalytic models of cosmic substructure \citep{2012NewA...17..175B,2021MNRAS.502..621J,2018PhRvD..97l3002H,2023MNRAS.523.1067S}.

        In this paper, we describe how \textit{future subhalos} -- halos that will fall into a more massive host -- differ from typical field halos even before infall. This pre-infall bias is part of every subhalo's history, distinct from the post-infall effects described above. It is a corollary of the \textit{assembly bias} phenomenon: at fixed halo mass, older halos (those that grew to a set fraction of their present mass earlier) tend to be more densely clustered \citep{2005MNRAS.363L..66G,2006ApJ...652...71W,2007MNRAS.377L...5G,2008ApJ...687...12D}. Assembly bias also runs the other way: halos in dense environments tend to be older than average halos of the same mass \citep{2004MNRAS.350.1385S,2006MNRAS.367.1039H,2007ApJ...654...53M}. The bias extends to halo density profiles as well, because older halos are more centrally concentrated \citep{1997ApJ...490..493N,2002ApJ...568...52W,2013MNRAS.432.1103L}. The effects of assembly bias are typically attributed to tidal forces from massive neighbors and from larger-scale structure, which suppress the growth of halos in dense regions \citep{2007MNRAS.375..633W,2009MNRAS.398.1742H,2011MNRAS.413.1973W,2015MNRAS.452.1958H,2018MNRAS.475.4411S,2018MNRAS.476.3631P,2020MNRAS.493.4763M}. Future subhalos are the extreme case, with the massive neighbor close enough to soon become the host. Indeed, \citet{2014ApJ...787..156B} found that infalling halos cease accreting and begin losing mass well outside the eventual host's virial radius, and \citet{2018ApJ...857..127S} reported qualitative differences between the assembly histories of infalling and field halos.
        
        Here we systematically characterize the pre-infall bias by measuring the progenitor mass functions (PMFs) of future subhalos in cosmological simulations. To ensure the bias is truly pre-infall, we restrict the measurement to halos that have never previously been subhalos, since a substantial fraction of halos near a massive host have already had a subhalo episode \citep{2021MNRAS.501.5948B}. We then develop a simple analytic model that captures the bias across cosmologies and infall epochs, and we use it to show how the bias translates into higher concentrations for subhalos near infall.
        
\begin{figure}
			\centering
			\includegraphics[width=\columnwidth,trim=8.9cm 5.75cm 9.65cm 4.95cm,clip]{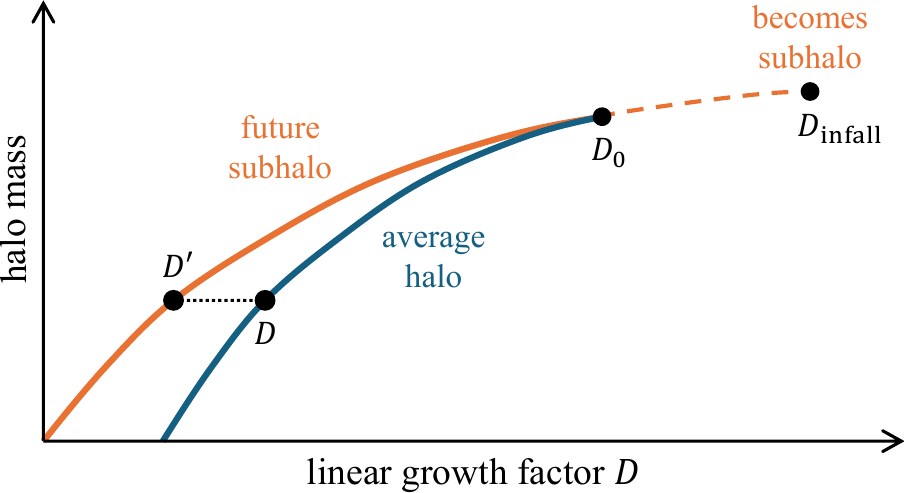}
			\caption{
            Illustration of the subhalo bias. The linear growth factor is a time parameter.
            A subset of halos at growth factor $D_0$ will become subhalos at $D_\mathrm{infall}$, and these \textit{future subhalos} have biased histories of past growth.
            The bias is captured by equation~(\ref{bias_intro}), which gives the relationship between $D'$ and $D$.
            }
			\label{fig:diagram}
		\end{figure}

        We find that the pre-infall bias can be expressed in the following compact form. Consider halos at scale factor $a_0$ and linear growth factor $D_0$, a subset of which will become subhalos when the growth factor is $D_\mathrm{infall}$. These future subhalos have progenitors at $a'<a_0$ and $D'<D_0$ that match average halos' progenitors at the growth factor $D$, where
        \begin{align}\label{bias_intro}
            \frac{\beta(D'/D_\mathrm{infall},a')}{D'} -\frac{\beta(D_0/D_\mathrm{infall},a_0)}{D_0} =\frac{1}{D}-\frac{1}{D_0}
        \end{align}
        with the bias being characterized by the function
        \begin{align}\label{beta_intro}
        \beta(x,a)&\equiv
        \begin{cases}
            (1-x)^{1.2+0.14a},\quad\text{for}\ M_{200\mathrm{c}},\\
            (1-x)^{1.2+0.05a},\quad\text{for}\ M_{200\mathrm{m}}.
        \end{cases}
        \end{align}
        Figure~\ref{fig:diagram} illustrates the setup. The scale-factor dependence accounts for the influence of dark energy at late times. Equation~(\ref{bias_intro}) is self-consistent by construction: within extended Press-Schechter (EPS) theory, it corresponds to the collapse barrier $\dc$ being replaced by
        \begin{align}
            \dc'&=\beta(D/D_\mathrm{infall},a)\dc
        \end{align}
        for future subhalos.
        
        This article is organized as follows. Section~\ref{sec:scalefree} calibrates the bias against scale-free simulations and derives its EPS interpretation. Section~\ref{sec:LCDM} extends the calibration to a $\Lambda$CDM cosmology, where dark energy modifies the model through the explicit scale-factor dependence in equation~(\ref{beta_intro}). Section~\ref{sec:c} applies the model to predict subhalo concentrations near the time of infall. We conclude in section~\ref{sec:conclusion}.
        
		\section{Characterizing the subhalo bias in scale-free simulations}\label{sec:scalefree}

        We analyze publicly available halo catalogues from the Erebos simulation suite \citep{2020ApJS..251...17D}. We primarily make use of the scale-free simulations of \citet{2015ApJ...799..108D}. These simulations assume matter domination and are initialized with power-law power spectra, $P(k)\propto k^n$. There are four simulations, corresponding to $n=-2.5$, $-2$, $-1.5$, and $-1$. Each simulation employs $1024^3$ particles. Halos from these simulations are identified with the \textsc{rockstar} halo finder \citep{2013ApJ...762..109B}, and the calculation of their masses is described by \citet{2020ApJS..251...17D}. We use the $M_{200}$ halo mass definition, which is the mass within the radius $R_{200}$ that encloses an average density 200 times the cosmological background density.
        
        Scale-free simulations are convenient because all times are equivalent, which allows all of the simulation snapshots to be stacked together to maximize halo sample sizes. In particular, at each scale factor $a$, we define a characteristic mass scale $M_*(a)$ such that $\sigma(M_*)=1$.
        Here
        \begin{align}\label{sigma}
            \sigma^2(M) = \int\frac{\diff^3\vec k}{(2\pi)^3}P(k) W^2(k R),
        \end{align}
        where $W(x)=(3/x^3)(\sin x-x\cos x)$ is the top-hat window function in Fourier space and $M=(4\pi/3)R^3\bar\rho$; here $\bar\rho$ is the mean matter density.
        In units of $M_*$, self-similarity of the initial conditions dictates that the halo distribution at each time is simply a different realization of the same underlying distribution.
        Hence, our general procedure for these simulations is to pick a range of nominal halo masses $M/M_*$, select halos from every snapshot that lie close to this nominal mass, and analyze the past assembly history of these halos.
        Note that simulation snapshots are separated by a factor of about 1.031 in the scale factor.

        At each snapshot scale factor $a_0$, we identify two classes of halo:
        \begin{itemize}
            \item \textbf{Field halos}: Halos that have never been subhalos at any $a\leq a_0$.
            \item \textbf{Future subhalos}: Halos that are never subhalos at any $a<a_\mathrm{infall}$ (with $a_\mathrm{infall}>a_0$), but are subhalos at $a=a_\mathrm{infall}$. We additionally restrict our selection to halos for which the new host is at least 10 times more massive than the subject halo (although we will test the influence of this restriction in section~\ref{sec:ratio}).
        \end{itemize}
        Here a subhalo is a halo that is inside a larger halo's $R_{200}$ radius.
        For future subhalos (class 2), we evaluate the mass ratio four snapshots before the subject becomes a subhalo ($a\simeq 0.90a_\mathrm{infall}$) to minimize the potential for the halos to pollute each other's $M_{200}$ masses. Also, in order to maximize sample sizes for distant-future subhalos, for each nominal value of $a_\mathrm{infall}/a_0$, we include halos whose infall times span a range of snapshots. Table~\ref{tab:infall} lists the $a_\mathrm{infall}/a_0$ and corresponding snapshot-number shifts that we use.

        \begin{table}
            \caption{Selection criteria for $a_\mathrm{infall}/a_0$.}\label{tab:infall}
            \begin{tabular*}{\columnwidth}{@{\extracolsep{\fill}} c c c }
		          \hline
                Nominal $a_\mathrm{infall}/a_0$ & $i_\mathrm{infall}-i_0$ & Actual $a_\mathrm{infall}/a_0$\\
		          \hline
                1.11 & 4 & 1.10--1.13 \\
                1.20 & 6--7 & 1.16--1.24 \\
                1.32 & 9--10 & 1.28--1.36 \\
                1.51 & 13--15 & 1.44--1.58 \\
                1.70 & 17--19 & 1.63--1.78 \\
                2.01 & 22--25 & 1.90--2.14 \\
            \end{tabular*}
            \tablecomments{The nominal $a_\mathrm{infall}/a_0$ is the logarithmic mean over the range of actual values. $i_\mathrm{infall}$ is the snapshot number in which a halo is first seen to be a subhalo (so $a_\mathrm{infall}$ lies between the $i_\mathrm{infall}-1$ and $i_\mathrm{infall}$ snapshots), while $i_0$ is the snapshot number corresponding to $a_0$.}
        \end{table}

        For each $a_0$ (each simulation snapshot), we identify field halos and future subhalos as described above.
        We then group these halos across all $a_0$ into bins in normalized mass $M(a_0)/M_*(a_0)$, using bins of width $\Delta\log_{10}M=0.05$. We consider only halos with mass $M_{200}>L=500m_\mathrm{p}$, where $m_\mathrm{p}$ is the simulation particle mass.

        \subsection{Main-progenitor mass accretion histories}\label{sec:growth}
        
		\begin{figure*}
			\centering
			\includegraphics[width=\linewidth]{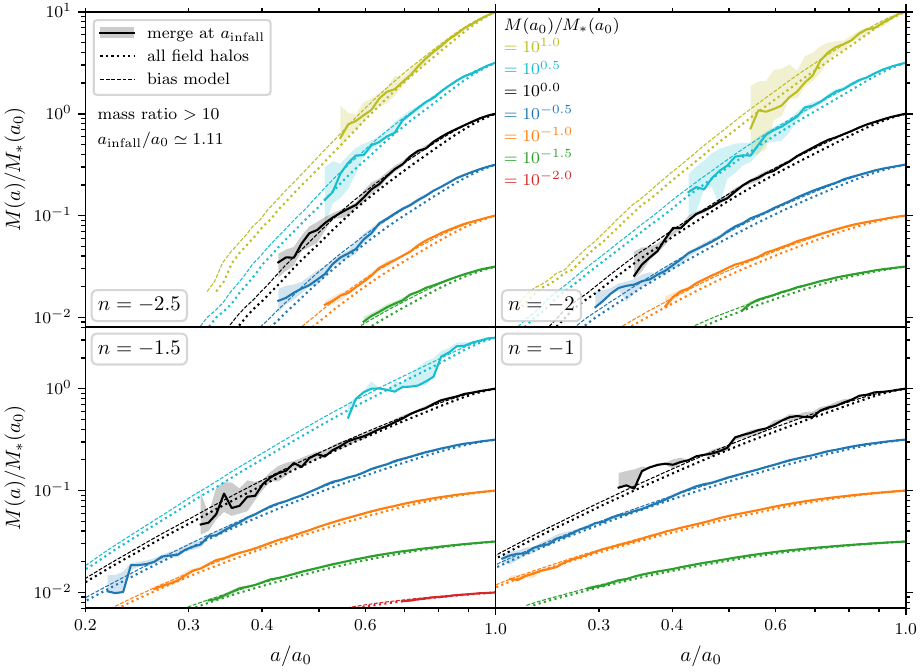}
			\caption{
            Main-progenitor growth histories for halos in a range of scale-free cosmologies (different panels). We select halos according to their mass at a final scale factor $a_0$ (in units of the characteristic nonlinear mass $M_*$), and we show the median main-progenitor mass at earlier scale factors $a<a_0$.
            Different colors represent different final masses $M(a_0)$.
            The dotted curves include all field halos, while the solid curves are restricted to field halos that will infall onto larger halos (becoming subhalos) at $a\simeq 1.1a_0$.
            These future subhalos are systematically biased to have slower growth histories.
            The dashed curves are growth histories predicted by the subhalo bias model in section~\ref{sec:EPS}, which takes only the (unbiased) field halo growth history as input.
            }
			\label{fig:growth}
		\end{figure*}

        For each of the four scale-free simulations, figure~\ref{fig:growth} shows the median main-progenitor growth histories for a range of $M(a_0)/M_*(a_0)$ (descendant masses).
        For each curve, we stack 7 adjacent mass bins (so we include halos within 0.175 dex of the nominal mass), but to avoid bias arising from the underlying descendant halo distribution, we weight descendant halos such that all mass bins contribute equally.
        At each nominal descendant mass and at each $a/a_0<1$, we evaluate the median value of $M(a)/M(a_0)$, the progenitor mass in units of the descendant mass. Note that in $M_*(a_0)$ units, each $a_0$ is associated with a different mass resolution limit $L$.
        We account for the inhomogeneous resolution limits by using the product limit estimator of \citet{KaplanMeier} to evaluate the median. For visual clarity, in figure~\ref{fig:growth} we multiply the median $M(a)/M(a_0)$ by the bin-center $M(a_0)$, so the vertical axis represents the growth history in $M_*$ units.
        The statistical error is essentially negligible for the field halos, but it is of concern for the future subhalos due to their much lower sample sizes. Thus, for the future subhalos, we additionally use the method of \citet{BrookmeyerCrowley} to estimate a 68 percent confidence interval on the median.

        The nature of the subhalo bias is evident in figure~\ref{fig:growth}. The future subhalos, with $a_\mathrm{infall}\simeq 1.11 a_0$, tend to be ``older'' than field halos of the same mass, in the sense that they grew more slowly in the past. At any fixed progenitor time $a/a_0<1$, the main progenitor of a future subhalo at $a_0$ tends to be more massive than that of a general field halo of the same mass at the same time. This progenitor mass difference tends to be higher for the cosmologies with more negative spectral indices $n$, and within a cosmology, it is higher for more massive descendant halos at $a_0$. Generally, when framed in terms of the progenitor masses at a fixed time, the subhalo bias is stronger for halos that assembled more rapidly.

        \subsection{Progenitor mass functions}\label{sec:cmf}
        
		\begin{figure*}
			\centering
			\includegraphics[width=\linewidth]{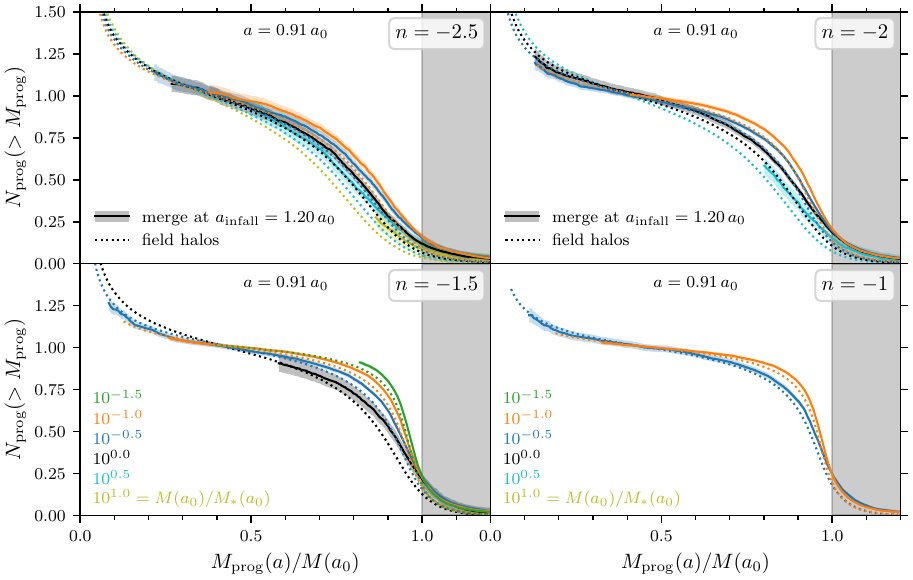}
			\caption{
            PMFs for halos in a range of scale-free cosmologies (different panels).
            We select halos according to their mass at a final scale factor $a_0$, and we show at $a\simeq 0.91a_0$ the cumulative number $N_\mathrm{prog}(>M_\mathrm{prog})$ of progenitor halos above the progenitor-to-descendant mass ratio $M_\mathrm{prog}(a)/M(a_0)$.
            Different colors represent different descendant halo masses $M(a_0)$.
            The dotted curves include all field halos, while the solid curves are restricted to field halos that will infall onto larger halos at $a\simeq 1.2a_0$.
            The progenitors of these future subhalos are biased toward higher masses, especially for very negative spectral indices $n$.
            The gray shaded region corresponds to progenitors of higher mass than the descendants.
            }
			\label{fig:cmf}
		\end{figure*}

        We next examine the full PMFs.
        For each descendant halo (at each $a_0$), we include all field halos at $a<a_0$ that are progenitors of either the subject halo or any of its subhalos. Progenitors are then binned based on their mass relative to their respective descendant halos, and the cumulative PMF $N_\mathrm{prog}(>M_\mathrm{prog})$ is obtained as the sum $\sum_i N_{\mathrm{prog},i}/N_{\mathrm{desc},i}$ over appropriate mass bins $i$, where $N_{\mathrm{prog},i}$ is the progenitor count in bin $i$ and $N_{\mathrm{desc},i}$ is the number of descendants that could have contributed progenitors to the mass bin (based on the $500m_\mathrm{p}$ resolution limit).
        We estimate the variance in $N_{\mathrm{prog},i}$ as $N_{\mathrm{prog},i}+1/2$ \citep[corresponding to a Jeffreys prior;][]{Jeffreys} and use this to estimate the uncertainty in $N_\mathrm{prog}(>M_\mathrm{prog})$.

        Figure~\ref{fig:cmf} shows the PMFs for the four scale-free cosmologies. Here we fix the progenitor epoch to be $a=0.91a_0$. We compare field halos to future subhalos, where for the future subhalos, we fix $a_\mathrm{infall}=1.20a_0$. We only show an uncertainty band for the future subhalos since uncertainty in the field halos is negligible in comparison.
        We show a range of nominal descendant halo masses, each of which includes halos within 0.275 dex of the nominal value. Specifically, we continue to use the descendant mass bins of width $\Delta\log_{10}M(a_0)=0.05$ discussed above, but we average the PMF over 11 consecutive bins, weighting each bin equally.
     	To suppress clutter, we truncate the future-subhalo PMFs when the uncertainty in $N_\mathrm{prog}$ exceeds 0.05, and we only show PMFs if they are nonzero and untruncated over a mass interval of at least 20\% of the descendant mass. When a future-subhalo PMF is omitted for this reason, we also omit the field-halo PMF for the same descendant mass (same color).

        Compared to the PMFs of field halos at large, those of future subhalos are systematically shifted to higher masses, with the shift being larger for more negative spectral index $n$. Note that less massive descendant halos (relative to $M_*$) are also associated with more massive progenitors (relative to the descendant mass). Remarkably, for the $n=-2$ case, the combination of these effects means that the PMF of a future subhalo precisely matches that of a general field halo that is half a decade less massive. That is, in the $n=-2$ panel of figure~\ref{fig:cmf}, the solid curves of each color overlap the dotted curves of the previous color. This observation motivates the more detailed analysis in section~\ref{sec:massshift}.
        
		\begin{figure*}
			\centering
			\includegraphics[width=\linewidth]{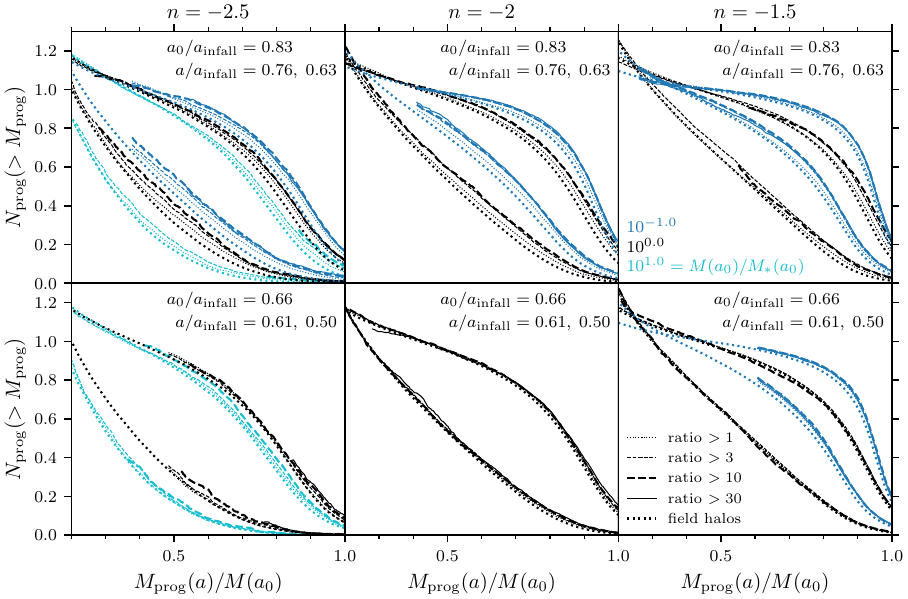}
			\caption{
            Dependence of PMFs of future subhalos on the mass ratio of the merger in which they become a subhalo.
            Like figure~\ref{fig:cmf}, we select halos according to their mass at a final scale factor $a_0$ (different colors), and we show at $a<a_0$ the cumulative number of progenitors of mass greater than $M_\mathrm{prog}$.
            The thick dotted curves show PMFs for field halos at large, while the other curves represent halos that merge onto a larger host halo at $a_\mathrm{infall}>a_0$ with a mass ratio larger than a threshold value; different line styles correspond to different thresholds.
            Across columns we vary the cosmology, while the two rows represent different merger times $a_\mathrm{infall}/a_0$. Each panel includes PMFs at two progenitor times $a/a_0$.
            }
			\label{fig:ratio}
		\end{figure*}

        We note that the PMFs have some support at $M_\mathrm{prog}/M(a_0)>1$ (gray shaded region). Such a feature is possible because of the halo mass definition. The $M_{200}$ definition does not generally encompass all material that fell into the halo, so it is possible for halo masses to temporarily decrease if enough recently accreted material crosses outside of $R_{200}$. More problematically, it is also possible for $M_{200}$ to be contaminated by material that belongs to another halo, especially if the nearby halo is much more massive than the subject halo. This consideration can be of special concern for future subhalos, since for $a$ close to $a_\mathrm{infall}$, these halos are guaranteed to have a much more massive halo nearby.
        It will present a potential source of systematic error for the analysis in section~\ref{sec:massshift}.

        \subsection{The merger mass ratio}\label{sec:ratio}
		
		For most of this work, we only consider future subhalos for which the eventual merger at $a_\mathrm{infall}$ is with a host halo at least 10 times as massive. We now test the influence of the threshold host-to-subhalo mass ratio. In figure~\ref{fig:ratio}, we compare PMFs of field halos to those of future subhalos with a range of mass-ratio thresholds. We consider a range of scale-free cosmologies, two different infall times $a_\mathrm{infall}/a_0$, two different progenitor times $a/a_0$, and several descendant halo masses $M(a_0)$. The PMFs are constructed and plotted as in section~\ref{sec:cmf}, aside from the choices above.
		
		Figure~\ref{fig:ratio} demonstrates that the subhalo bias is converged at a mass ratio of 10 (thick dashed curves). Increasing the threshold mass ratio to 30 (thin solid curves, only shown in a few of the panels) does not shift the future-subhalo PMFs.
		Often, lower-threshold PMFs (thin dotted or dashed curves) are also at the converged value, but this is not always the case. In general, the future-subhalo PMFs with threshold mass ratios of 1 or 3 lie between the threshold-10 PMFs and the general field halo PMFs. These findings motivate our choice to adopt a mass ratio threshold of 10 for the remainder of this work.
        
        \subsection{Subhalo bias as an effective mass shift}\label{sec:massshift}
        
        \begin{figure*}
        	\centering
        	\includegraphics[width=\linewidth]{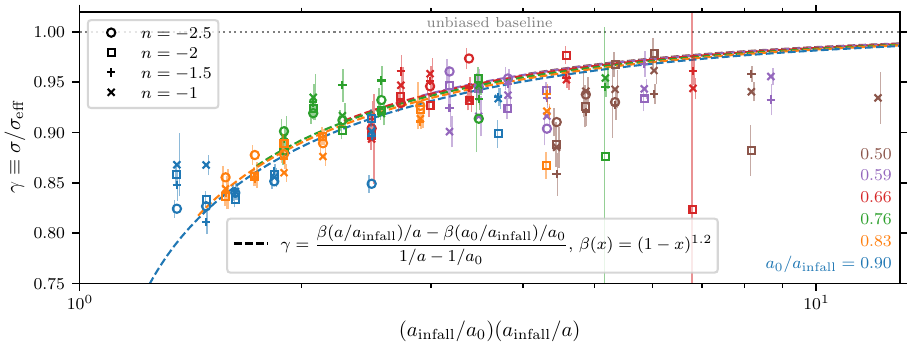}
        	\caption{
        		In the scale-free cosmologies, PMFs of future subhalos of mass $M$ match PMFs of average halos of mass $M_\mathrm{eff}$. For halos at scale factor $a_0$, we plot the bias parameter $\gamma=\sigma/\sigma_\mathrm{eff}=\sigma(M)/\sigma(M_\mathrm{eff})$ for different cosmologies (different markers), different progenitor times $a/a_0$, and different infall times $a_\mathrm{infall}/a_0$ (colors).
        		The dashed curves show the bias model in section~\ref{sec:EPS} for each $a_\mathrm{infall}/a_0$.
                Points of the same color (infall time) at different horizontal positions have different progenitor times $a/a_0$.
        	}
        	\label{fig:massbias}
        \end{figure*}
        
        We now quantify how the PMFs of future subhalos are biased. To motivate this analysis, notice how in figure~\ref{fig:cmf}, the descendant masses $M(a_0)$ are spaced in intervals of 0.5 dex, corresponding to a factor of 3.2.
        For $n=-2$, the PMFs of future subhalos are shifted by one descendant-mass interval, meaning that PMFs of future subhalos of mass $M$ match PMFs of average halos of mass $M/3.2$, when considered in units of the descendant mass.
        For $n=-2.5$, the PMFs of future subhalos are shifted instead by about two descendant-mass intervals, while for $n=-1$, the shift is closer to half of a descendant-mass interval.
        But for these cosmologies, $\diff\ln \sigma/\diff\ln M=-(n+3)/6$, so all of these shifts correspond to about the same factor of $\sigma$.
        That is, PMFs of future subhalos of mass $M$ approximately match PMFs of average halos of mass $M_\mathrm{eff}$ such that $\sigma(M_\mathrm{eff})=\sigma(M)/\gamma$, where for the configurations in figure~\ref{fig:cmf}, $\gamma\simeq 0.83$ uniformly across cosmologies.
        
        We now measure the bias parameter $\gamma$ systematically. We consider the 6 progenitor times $a/a_0=0.91$, 0.83, 0.76, 0.67, 0.50, and 0.33. For each cosmology (i.e., spectral index $n$), each future-subhalo infall time $a_\mathrm{infall}/a_0$ (table~\ref{tab:infall}), and each progenitor time $a/a_0$, we calculate an optimal $M_\mathrm{eff}/M$ such that PMFs of future subhalos of mass $M$ match PMFs of average halos of mass $M_\mathrm{eff}$. We then evaluate $\gamma=(M_\mathrm{eff}/M)^{(n+3)/6}$. The results are shown in figure~\ref{fig:massbias}.
        
        The optimal $M_\mathrm{eff}/M$ is obtained by finding the root of a weighted sum over the differences $d_j$ between the mean $M_\mathrm{prog}$ for future subhalos and the mean $M_\mathrm{prog}$ for average halos, with both considered in the $j$-th descendant mass bin (of width 0.05 dex). For the evaluation of $d_j$, several steps are taken to mitigate systematic error. Each evaluation of the mean progenitor mass is made between $m_1$ and $m_2$, where $N_\mathrm{prog}(>m_1)=N_\mathrm{max}$ and $N_\mathrm{prog}(>m_2)=N_\mathrm{min}$, with the same $N_\mathrm{max}$ and $N_\mathrm{min}$ used for the future subhalos as for the halos at large. We take $N_\mathrm{max}=\min(1,N'_\mathrm{max})$, where $N'_\mathrm{max}$ is the highest $N_\mathrm{prog}$ value at which both sides of the comparison remain above the resolution limit. Meanwhile, we set $N_\mathrm{min}=0.2+N'_\mathrm{min}$, where $N'_\mathrm{min}=N_\mathrm{prog}[>M(a_0)]$ is the number of progenitors more massive than the descendant. As we discussed in section~\ref{sec:cmf}, even before infall, the mass of a future subhalo could include mass that is bound to its future host halo. The $N_\mathrm{min}$ truncation is intended to minimize the influence of this effect. One difficulty, however, is that $N_\mathrm{min}$ and $N_\mathrm{max}$ are biased if the progenitor distribution is skewed at their locations. For the mean mass between $N_\mathrm{min}$ and $N_\mathrm{max}$, this bias is amplified if the mean is evaluated using the same data as $N_\mathrm{min}$ and $N_\mathrm{max}$. To suppress this effect, we partition the simulation snapshots into blocks of length $\Delta\log a \simeq [(n+3)/2] \Delta\log M$ (where $\Delta\log M$ is the 0.05 dex bin width), and we split the future subhalos into two sets, one drawn from the even-numbered blocks and the other from the odd-numbered blocks. This $\Delta\log a$ is set to suppress correlations that would result from halos contributing to the same mass bin in different blocks. We use one set of descendants to find $N_\mathrm{min}$ and $N_\mathrm{max}$ and the other to evaluate the mean progenitor mass and hence $d_j$; we switch the roles of the sets and evaluate $d_j$ again; and we average the results. We only use this procedure for the future subhalos, because the overall field-halo set has rich enough statistics that errors are negligible.
        
        Uncertainties are estimated by assuming Poisson errors in the PMFs, which are appropriately propagated into uncertainty in the mean mass (and hence $d_j$) evaluation, both directly and through uncertainty in the boundaries $m_1$ and $m_2$. We only consider uncertainty on the future-subhalo side of the $d_j$ evaluation; uncertainty on the all-halos side is negligible in comparison.
        We then evaluate the sum over $d_j$ with inverse-variance weighting to minimize error. Specifically, we evaluate $d=\sum_j d_j/\sigma_{d_j}^2$ and find its root, where $\sigma_{d_j}$ is the uncertainty in $d_j$. In general, errors are underestimated due to correlations in the data, so each $\sigma_{d_j}^2$ includes a uniformly added variance term calibrated so that the $d_j/\sigma_{d_j}$ have variance 1 \citep[this corresponds to random-effects modeling as in][]{1982NISTJ..87..377P}.
        Finally, the error in the root value of $M_\mathrm{eff}/M$ is obtained from the slope of the $d$--$M_\mathrm{eff}/M$ curve at the root.
        
        Figure~\ref{fig:massbias} shows that the bias parameters $\gamma$ obtained from this procedure generally support the idea that, for scale-free cosmologies, the subhalo bias is captured by a $\sigma(M_\mathrm{eff})=\sigma(M)/\gamma$ model. At fixed infall time $a_\mathrm{infall}/a_0$ and fixed progenitor time $a/a_0$, the different cosmologies require approximately the same $\gamma$, although the $\gamma$ can become noisy at low $a/a_0$ and high $a_\mathrm{infall}/a_0$. Note that this is the regime where there are few future subhalos.
        
        Interestingly, figure~\ref{fig:massbias} suggests that $\gamma$ may depend on the progenitor and infall times only through the combination $(a_\mathrm{infall}/a_0)(a_\mathrm{infall}/a)$.
        However, it turns out that this particular time dependence cannot yield a self-consistent model. We discuss next how to construct a self-consistent subhalo bias model using these measured bias parameters $\gamma$.
        
        \subsection{EPS interpretation of the subhalo bias}\label{sec:EPS}
        
		
		We now show how the subhalo bias, which we framed as an effective shift in the descendant mass, can be naturally interpreted within the context of EPS theory.
		For future clarity, we will discuss EPS theory in terms of the linear growth factor $D=D(a)$, rather than the scale factor $a$. However, in the scale-free simulations, $D(a)=a$.

		Consider a descendant halo of mass $M_0$, with $\sigma_0=\sigma(M_0)$, at a time when the linear growth factor is $D_0$, and consider progenitors of mass $M$, with $\sigma=\sigma(M)$, when the linear growth factor is $D$. Let $\delta_0=\dc/D_0$ and $\delta=\dc/D$, where $\dc$ is the collapse barrier in EPS theory. Then the EPS PMF is
        \begin{align}
        \frac{\diff N}{\diff \ln M}
        &=
        \sqrt{\frac{2}{\pi}}\frac{M_0}{M}\sigma^2
        \frac{\delta-\delta_0}{(\sigma^2-\sigma^2_0)^{3/2}}
        \nonumber\\&\hphantom{=}\times
        \exp\!\left[-\frac{(\delta-\delta_0)^2}{2(\sigma^2-\sigma^2_0)}\right]
        \left|\frac{\diff \ln\sigma}{\diff\ln M}\right|
        \end{align}
        For a scale-free cosmology with $P(k)\propto k^n$, $\sigma(M)\propto M^{-(3+n)/6}$. Hence the PMF becomes
        \begin{align}
        	\frac{\diff N}{\diff \ln M}
        	&=
        	\sqrt{\frac{2}{\pi}}
        	\frac{3+n}{6}
        	\frac{(M/M_0)^{-(4+n)/3}}{[(M/M_0)^{-(3+n)/3}-1]^{3/2}}
        	\frac{\delta-\delta_0}{\sigma_0}
        	\nonumber\\&\hphantom{=}\times
        	\exp\!\left[-\frac{1}{2[(M/M_0)^{-(3+n)/3}-1]}\left(\frac{\delta-\delta_0}{\sigma_0}\right)^2\right].
        \end{align}
        Notably, at fixed progenitor mass fraction $M/M_0$, the PMF depends only on the ratio $(\delta-\delta_0)/\sigma_0$.
        
        In section~\ref{sec:massshift}, we found that progenitors of future subhalos with descendant mass corresponding to $\sigma_0$ match progenitors of general field halos with mass corresponding to $\sigma_0/\gamma$.
        Since the PMF depends only on $(\delta-\delta_0)/\sigma_0=(1/D-1/D_0)\dc/\sigma_0$, we can equivalently say that $1/D-1/D_0$ for future subhalos matches $\gamma(1/D-1/D_0)$ for general field halos.
        More precisely,
        progenitors at growth factor $D'$ of future subhalos match progenitors at $D$ of general field halos, with 
        \begin{align}\label{gamma_D}
        	1/D'-1/D_0= \gamma (1/D-1/D_0).
        \end{align}
        That is, at fixed progenitor mass fraction, the subhalo bias (with $\gamma<1$) corresponds to shifting the progenitor to an earlier time by an amount set by $\gamma$.
        
        We can use this correspondence to impose a consistency requirement on how $\gamma$ depends on the times $D$ and $D_0$. The requirement is that each of the two terms of the expression $\gamma (1/D-1/D_0)$ depend on $D$ or $D_0$ alone.
        This condition corresponds to
        \begin{align}\label{gamma}
        	\gamma(D,D_0) (1/D-1/D_0) &=  \beta(D)/D-\beta(D_0)/D_0
        \end{align}
        for some function $\beta$.
        For example, this is required to ensure that the PMF at $D$ of a descendant at $D_1$, convolved with the PMF at $D_1$ of a descendant at $D_0$, matches the PMF at $D$ of a descendant at $D_0$.
        
        To motivate a functional form for $\beta$, consider a halo that merges onto a larger halo at growth factor $D_\mathrm{infall}$. In the EPS context, at earlier times $D<D_\mathrm{infall}$, the subject halo is in a large-scale overdensity $\dc D/D_\mathrm{infall}$. Consequently, the effective collapse threshold is reduced by this amount, becoming $(1-D/D_\mathrm{infall})\dc$.
        This picture suggests $\beta(D)=1-D/D_\mathrm{infall}$. As is well known, this form unfortunately does not produce the needed bias; the $D$-dependent part cancels between the two terms in equation~(\ref{gamma}).\footnote{In the language of excursion set theory \citep{1991ApJ...379..440B}, the absence of assembly history bias is a consequence of using uncorrelated steps in the EPS formalism. However, we show in appendix~\ref{sec:excursionset} that correctly accounting for the correlations implied by the top-hat window function in equation~(\ref{sigma}) leads to predicting bias in the wrong direction \citep[see also][section IX]{2007IJMPD..16..763Z}.} However, we can make a simple modification,
        \begin{align}\label{beta}
        	\beta(D/D_\mathrm{infall})
        	&=
        	(1-D/D_\mathrm{infall})^p
        \end{align}
        for some $p\neq 1$.
        
        We fit equation~(\ref{gamma}), with $\beta$ given by equation~(\ref{beta}), simultaneously to every point in figure~\ref{fig:massbias}. For this fit, we use the uncertainty in each $\gamma$, but we add to each point's variance a uniform value tuned so that the reduced $\chi^2$ of the fit is 1. This variance-addition procedure \citep[which corresponds to random-effects modeling as in][]{1982NISTJ..87..377P} is approximately justified because there are correlations across the different snapshots of the scale-free simulations, which lead to uncertainty being underestimated.
        
		\begin{figure*}
			\centering
			\includegraphics[width=\linewidth]{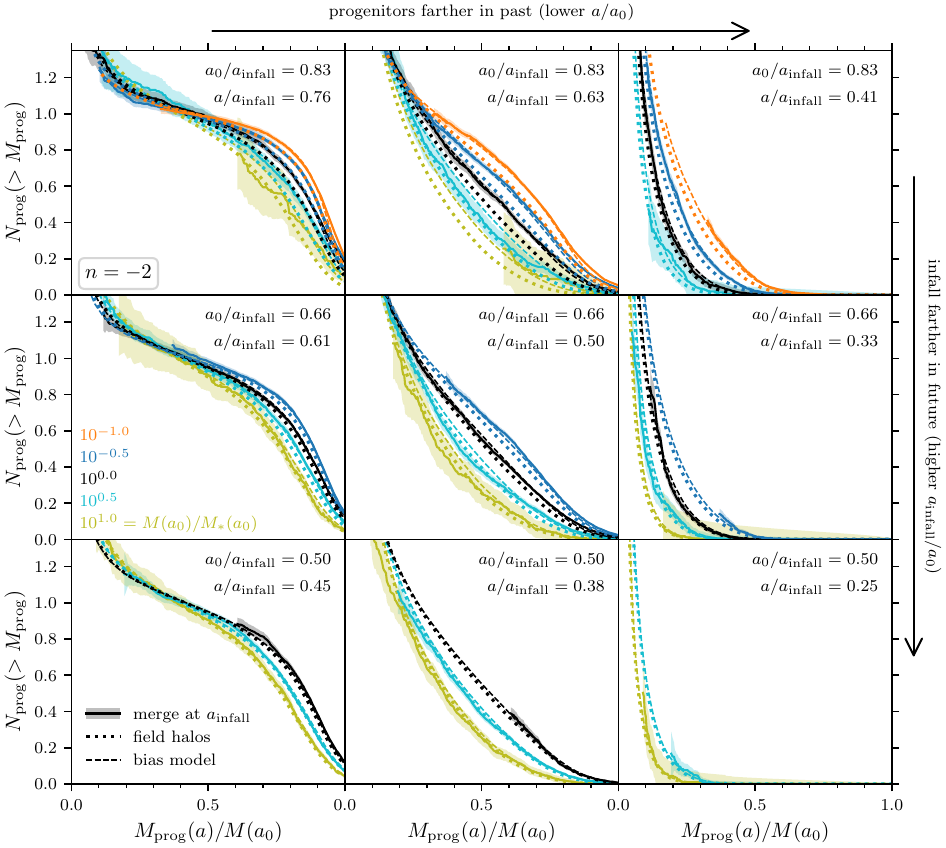}
			\caption{
			PMFs of halos in the $n=-2$ scale-free simulation.
            We select halos according to their mass at a final scale factor $a_0$, and we show the progenitor distribution at $a<a_0$, with different columns representing different $a/a_0$. Colors indicate the descendant mass; we include halos that lie within 0.025 dex of the nominal mass.
            The dotted curves include all field halos, while the solid curves are restricted to field halos that will infall onto larger halos at some scale factor $a_\mathrm{infall}>a_0$, and different rows show different $a_0/a_\mathrm{infall}$.
            With the thin dashed curves, we test the bias model in section~\ref{sec:EPS} with $p=1.20$; these curves show the progenitor distributions for field halos of mass lower by a factor of $\gamma^{6/(3+n)}$ (corresponding to $\sigma$ higher by a factor of $\gamma^{-1}$), with $\gamma$ given by equation~(\ref{gamma}).
            }
			\label{fig:cmfs1}
		\end{figure*}

		\begin{figure*}
			\centering
			\includegraphics[width=\linewidth]{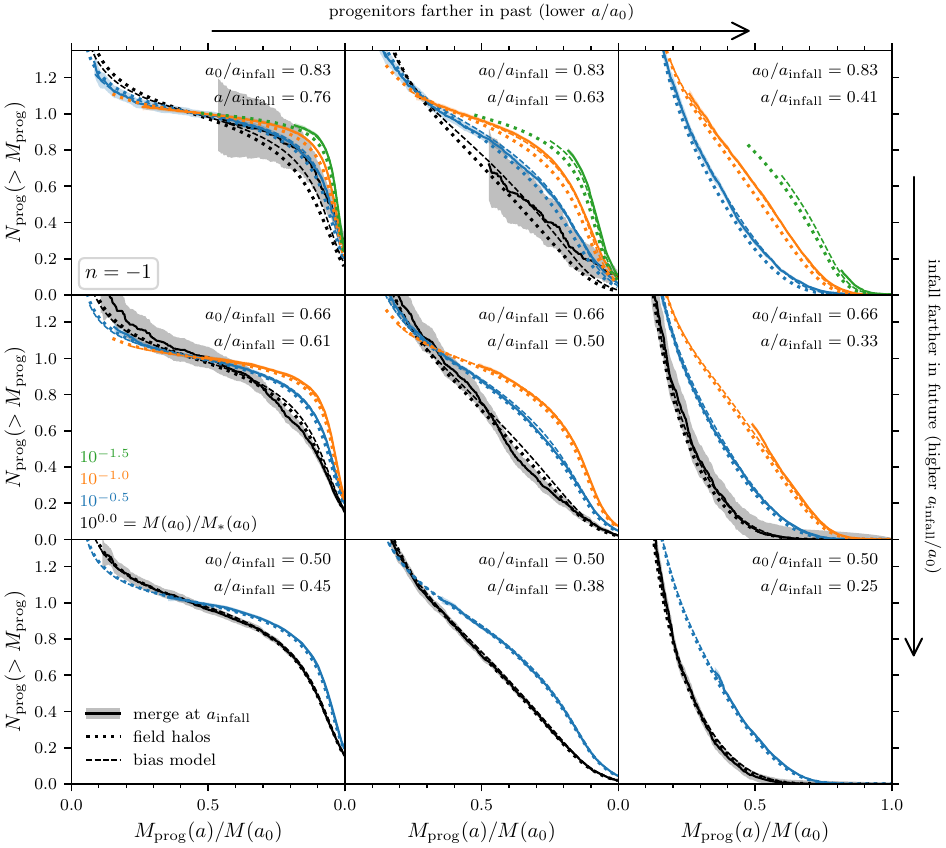}
			\caption{
            Like figure~\ref{fig:cmfs1} but for the $n=-1$ simulation.
            }
			\label{fig:cmfs2}
		\end{figure*}
        
        We obtain a best-fitting value of $p=1.20$. The corresponding $\gamma$ curves are shown (separately for each $a_0/a_\mathrm{infall}$) as dashed curves in figure~\ref{fig:massbias}.
        These curves agree reasonably well with the general trend of the measured $\gamma$.
        Significant deviations occur mainly in the following two regimes.
        \begin{enumerate}
        	\item For the highest $a_0/a_\mathrm{infall}$ (blue), $\gamma$ can deviate significantly from the fit. Note that in these cases, the merger occurs only shortly after the descendant time, so the descendant mass may be contaminated by the outskirts of the future host.
        	\item At low $a/a_\mathrm{infall}$ and low $a_0/a_\mathrm{infall}$, $\gamma$ tends to lie significantly below the fit (so the fit corresponds to less bias).
        	This offset is likely due to bias in the quantile boundaries between which the mean progenitor mass is evaluated, which then translates into bias in the mean progenitor mass. For small $a/a_\mathrm{infall}$ and $a_0/a_\mathrm{infall}$, there are few progenitors, and the raw distribution of progenitors above the $500m_\mathrm{p}$ resolution limit (before correcting for how many descendants could contribute progenitors) is strongly biased toward high masses. This skew would be expected to bias the quantiles upward in mass, corresponding to overestimated progenitor masses and hence overestimated subhalo bias. We have verified that the $\gamma$ values in this regime decrease further if we make the resolution limit more stringent by raising it to $1500m_\mathrm{p}$, while the $\gamma$ at higher $a/a_\mathrm{infall}$ and $a_0/a_\mathrm{infall}$ are unaffected.
        \end{enumerate}
        Importantly, we have reason to expect that both regimes suffer from systematic errors. We tried removing either or both of the $a_0/a_\mathrm{infall}>0.9$ (blue) and $a_0/a_\mathrm{infall}\leq 0.6$ (brown and purple) regimes from the fit, and the best-fitting $p=1.20$ did not shift by more than 0.01. We conclude that the value $p=1.20$ is robustly inferred from the simulation data.
        
        Figures \ref{fig:cmfs1} and~\ref{fig:cmfs2} show how the bias model with $p=1.20$ works for the PMFs in the $n=-2$ and $n=-1$ simulations, respectively. For a range of descendant masses, $a_\mathrm{infall}/a_0$, and $a/a_0$, we compare the progenitor distributions of the future subhalos (solid curves) and the general field halos (dotted curves). The dashed curves show the progenitor distributions of general field halos of mass lower by a factor of $\gamma^{6/(3+n)}$, corresponding to $\sigma$ higher by a factor of $\gamma^{-1}$, with $\gamma$ given by equation~(\ref{gamma}) and $\beta$ given by equation~(\ref{beta}). These shifted field-halo progenitor distributions agree well with the future-subhalo progenitor distributions (to such a degree that the dashed curves are often hidden underneath the solid curves), demonstrating the accuracy of the bias model.
		
		We can additionally test the EPS interpretation of the bias model directly.
		Let $D'$ be the growth factor in the biased case that corresponds to $D$ for the unbiased case. Then
		\begin{align}\label{growthbias0}
			\frac{\beta(D')}{D'}-\frac{\beta(D_0)}{D_0}
			&=
			\frac{1}{D}-\frac{1}{D_0}.
		\end{align}
		We now consider the median main-progenitor growth histories of the general field halos (dotted curves) in figure~\ref{fig:growth}. At each mass $M(a)$, we numerically solve equation~(\ref{growthbias0}) to find the bias-shifted time $a'$ (recall $D=a$ for these simulations, so $D'=a'$), and we plot in figure~\ref{fig:growth} the resulting $M(a')$ growth histories as dashed curves. These curves agree well with the future-subhalo growth histories (solid curves), providing further validation of the bias model.

        \section{Concordance cosmology: the influence of dark energy}\label{sec:LCDM}
        
		\begin{figure*}
			\centering
			\includegraphics[width=\linewidth]{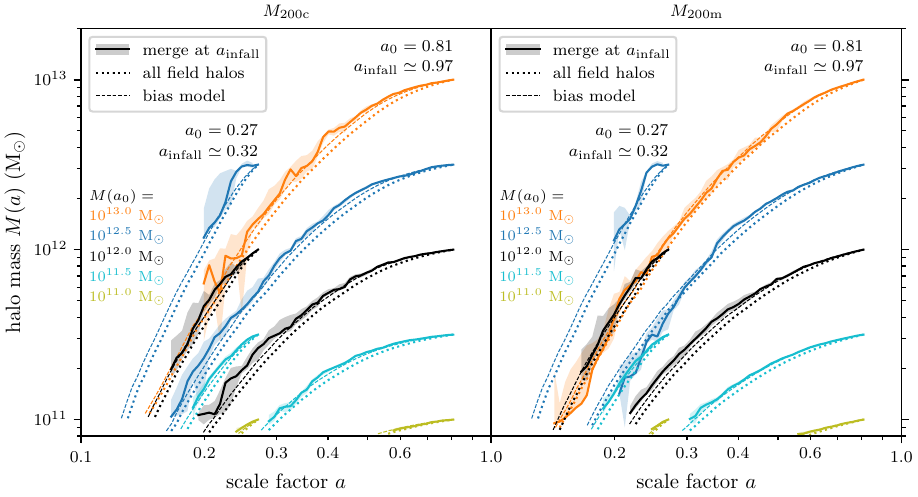}
			\caption{
            Main-progenitor growth histories for halos in a standard concordance cosmology.
            We consider two final scale factors $a_0$, and for each $a_0$ we select halos according to their mass at this time and show the median main-progenitor mass at earlier scale factors $a<a_0$.
            Different colors represent different final masses $M(a_0)$; we select halos that lie within 0.275 dex of the nominal mass.
            The dotted curves include all field halos, while the solid curves are restricted to field halos that will infall onto larger halos (becoming subhalos) at $a_\mathrm{infall}\simeq 1.2a_0$.
            These future subhalos are biased to have slower growth histories.
            The dashed curves show future-subhalo growth histories predicted by our bias model, which takes the field-halo growth histories as input.
            }
			\label{fig:growth-LCDM}
		\end{figure*}
        
        We now consider a standard concordance cosmology.
        We use the simulations from \citet{2015ApJ...799..108D}, which were carried out with \citet{2014A&A...571A..16P} cosmological parameters.
        Halo catalogues from these simulations are included in the same Erebos simulation suite \citep{2020ApJS..251...17D}. There are three simulations with periodic box sizes 187, 373, and 746~Mpc. We stack these simulations together to maximize sample sizes.
        
        Compared to the scale-free simulations, the initial power spectrum is not a pure power law, and the universe is not dominated by matter at all times. In particular, we must further specify the halo mass definition. We will consider both the $M_{200\mathrm{c}}$ and $M_{200\mathrm{m}}$ mass definitions. The former is the mass inside the radius $R_{200\mathrm{c}}$ enclosing 200 times the critical density, while the latter is the mass inside the radius $R_{200\mathrm{m}}$ enclosing 200 times the mean matter density. We also appropriately redefine subhalo infall to correspond to crossing the host halo's $R_{200\mathrm{c}}$ or $R_{200\mathrm{m}}$ radius, respectively.
        
        \subsection{Subhalo bias in the growth histories}\label{sec:growthLCDM}
        
        Figure~\ref{fig:growth-LCDM} shows the median main-progenitor growth histories for descendant halos at $a_0=0.32$ and $a_0=0.81$, evaluated using the same methods as in section~\ref{sec:growth}. We consider general field halos (dotted curves) along with future subhalos with $a_\mathrm{infall}\simeq 1.2a_0$ (solid curves). Similarly to what we noticed in section~\ref{sec:growth}, progenitors of future subhalos are biased upward in mass at fixed time, or equivalently they are biased toward earlier times at fixed progenitor mass.
		
		We now evaluate the bias parameter $\gamma$ from section~\ref{sec:massshift} for these simulations. Whereas previously we measured $\gamma$ from the effective shift in the descendant mass, now, based on the discussion in section~\ref{sec:EPS}, we measure $\gamma$ by comparing the field-halo and future-subhalo growth histories. At each main-progenitor mass $M(a)$ for the general field halos, we find the time $a'$ at which the main progenitors of future subhalos reach the same mass.
		Then from equation~(\ref{gamma_D}) we have
		\begin{align}\label{gamma_D_shift}
			\gamma
			&=
			\frac{1/D'-1/D_0}{1/D-1/D_0},
		\end{align}
		where as usual $D$, $D'$, and $D_0$ are the linear growth factors associated with $a$, $a'$, and $a_0$, respectively.

        We consider 40 values of the main-progenitor mass fraction $M(a)/M(a_0)$ ranging from 0.01 to 1; the five infall redshifts $z_\mathrm{infall}=0$, 1, 2, 3, and 4; and five of the six relative descendant times in table~\ref{tab:infall}. We omit $a_\mathrm{infall}/a_0=1.11$ because we find that, due to dark energy's influence on structure formation, the future subhalos' masses are often already significantly inflated by the host by this $a_0$ (especially for the $M_{200\mathrm{c}}$ mass definition).
        In each case, we evaluate $\gamma$ using equation~(\ref{gamma_D_shift}) for descendant halo masses $M(a_0)$ ranging from $10^{11}~\Msol$ upward in intervals of 0.35 dex. We continue to use bins of width 0.05 dex, but we consider groups of 7 consecutive mass bins, with halos weighted to ensure that each bin contributes equally to the group.
		We estimate the uncertainty in $\gamma$ by appropriately propagating the uncertainty in the growth history.
		Finally, we take the inverse-variance-weighted average value of $\gamma$ over all of the descendant mass groups.
        
		\begin{figure*}
			\centering
			\includegraphics[width=\textwidth]{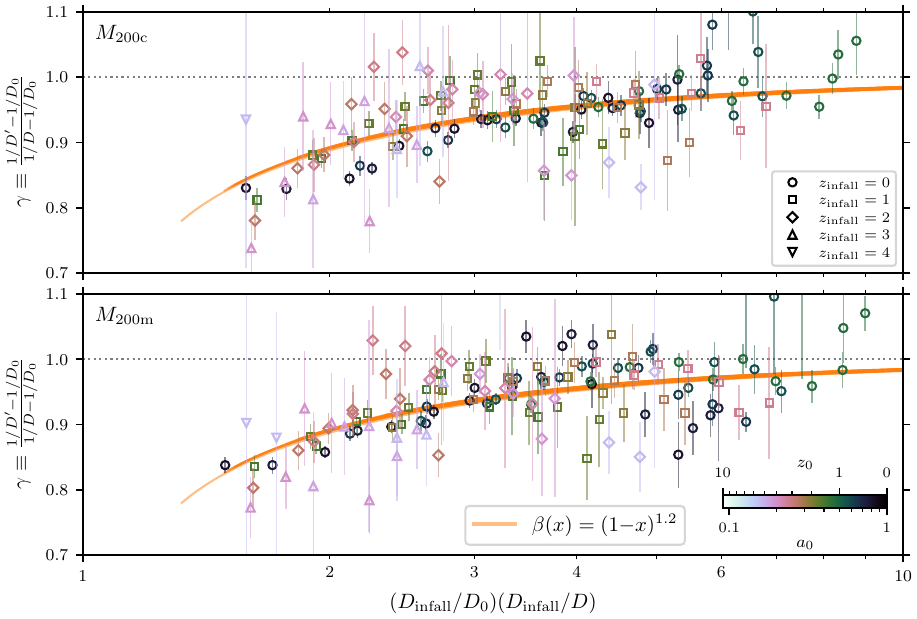}
			\caption{
            Bias parameters $\gamma$ measured using equation~(\ref{gamma_D_shift}) from the median main-progenitor growth histories in concordance cosmology simulations. Future subhalos at growth factor $D_0$ (becoming subhalos at $D_\mathrm{infall}$) have growth histories shifted earlier in time relative to average halos, such that they reached a mass at growth factor $D'$ that average halos reached at $D>D'$ (see figure~\ref{fig:diagram}).
            The upper panel considers the $M_{200\mathrm{c}}$ mass definition, while the lower panel considers $M_{200\mathrm{m}}$. Different markers represent different infall redshifts $z_\mathrm{infall}$, the color scale denotes descendant redshift $z_0$, and points of the same color at different horizontal positions have different progenitor times $D/D_0$.
            The orange curves have the same meaning as in figure~\ref{fig:massbias}: they represent the bias model of section~\ref{sec:EPS} before accounting for the effect of dark energy at late times.
            }
			\label{fig:shifts_LCDM}
		\end{figure*}
		
		Figure~\ref{fig:shifts_LCDM} shows the bias parameters $\gamma$ measured in this way, with results plotted separately for the $M_{200\mathrm{c}}$ and $M_{200\mathrm{m}}$ mass definitions. Here, to reduce clutter, we show only every third value of $M(a)/M(a_0)$. For comparison, we also show the range of $\gamma$ predicted by the bias prescription in section~\ref{sec:EPS} (solid curves) for the same configurations. The measured $\gamma$ follow the same trend as the predictions but with some offsets, as we discuss next.
		
		\subsection{Accounting for dark energy in the bias model}\label{sec:modelLCDM}

		The bias parameters $\gamma$ in figure~\ref{fig:shifts_LCDM} are colored based on the descendant redshift. At low redshifts (dark colors), the measured $\gamma$ lie systematically below the predictions from section~\ref{sec:EPS}, corresponding to more subhalo bias than expected. This offset is especially significant because the large number of descendant halos at low redshifts suppresses the uncertainty in the bias measurements. Note that the discrepancy is more pronounced for the $M_{200\mathrm{c}}$ mass definition than for the $M_{200\mathrm{m}}$ definition.

        The model from section~\ref{sec:EPS} was calibrated for scale-free simulations in a matter-dominated cosmology. At low redshifts, the influence of dark energy becomes important. To approximately model dark energy's influence, we make a simple modification to the bias model to include explicit redshift-dependence. Instead of equation~(\ref{beta}), we define $\beta$ as
        \begin{align}\label{beta_a}
        	\beta(D/D_\mathrm{infall},a)
        	&=
        	(1-D/D_\mathrm{infall})^{p+q\, a},
        \end{align}
        where $p=1.2$ as before but with a new parameter $q$, which parametrizes the redshift-dependence. Here $a$ is the scale factor.

        We fit $\gamma$ in the form of equation~(\ref{gamma_D}), with $\beta$ given by equation~(\ref{beta_a}), to the $\gamma$ measured in section~\ref{sec:growthLCDM}. Here we use all 40 of the $M(a)/M(a_0)$ values, rather than the thinned set shown in figure~\ref{fig:shifts_LCDM}. We obtain $q=0.140$ for the $M_{200\mathrm{c}}$ mass definition and $q=0.050$ for the $M_{200\mathrm{m}}$ mass definition. The uncertainties in $q$ are at the 0.01 level.
        In figure~\ref{fig:growth-LCDM}, the thin dashed curves show the main-progenitor growth histories of the future subhalos predicted by equation~(\ref{beta_a}) with these calibrations, based on the average-halo growth histories. These model growth histories match the measured future-subhalo growth histories well.

        \section{Implication for subhalo concentrations}\label{sec:c}

        We have shown that halos that will become subhalos in the future have biased assembly histories. But a halo's assembly history sets the radial profile of its internal density; halos that grew more slowly are more centrally concentrated in mass \citep[e.g.][]{1997ApJ...490..493N,2002ApJ...568...52W,2010arXiv1010.2539D,2013MNRAS.432.1103L,2019PhRvD.100b3523D}.
        Consequently, future subhalos are expected to be more centrally concentrated than field halos of the same mass at the same time. Indeed, \citet[appendix A]{2025arXiv251204156D} noticed in the simulations of \citet{2023MNRAS.518.3509D} that future subhalos are 10--20\% more concentrated than average field halos of the same mass shortly before they become subhalos.

        We now quantify the future-subhalo concentration bias using the halo concentration model of \citet{2016MNRAS.460.1214L}.
        We adopt \citet{1965TrAlm...5...87E} density profiles \citep{2004MNRAS.349.1039N},        
        \begin{equation}\label{einasto}
        	\rho(r)\propto\exp\left[-\frac{2}{\alpha}\left(\frac{r}{r_{-2}}\right)^\alpha\right],
        \end{equation}
        with $\alpha=0.18$. Here $r_{-2}$ is the radius at which the logarithmic density slope is $\diff\ln\rho/\diff\ln r=-2$, and it is common to parametrize this in terms of the concentration parameter $c=R_\mathrm{vir}/r_{-2}$, where $R_\mathrm{vir}$ is the virial radius. The overall normalization of equation~(\ref{einasto}) is set by the halo virial mass $M_\mathrm{vir}$. For consistency with \citet{2016MNRAS.460.1214L}, we use the $M_{200\mathrm{c}}$ and $R_{200\mathrm{c}}$ definition.

\begin{figure}
			\centering
			\includegraphics[width=\columnwidth]{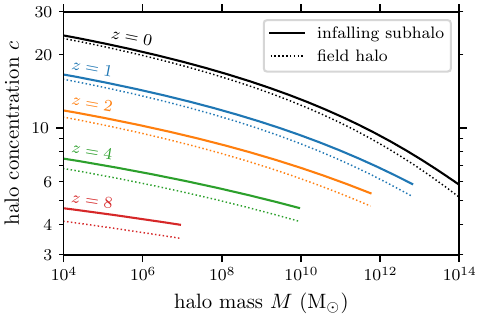}
			\caption{
            Bias in subhalo concentration parameters. Dotted curves show the concentration-mass relation of \citet{2016MNRAS.460.1214L} at several different redshifts (different colors), while the solid curves show halo concentration parameters modified according to the subhalo assembly bias model for $a=a_\mathrm{infall}$. Subhalos at the time of infall have 10--15\% higher concentration parameters than field halos.
            }
			\label{fig:c}
		\end{figure}

        The model of \citet{2016MNRAS.460.1214L} considers the mass $M_{-2}$ enclosed within $r_{-2}$.
        According to the model, the average enclosed density satisfies
        \begin{align}\label{ludlow1}
            \frac{M_{-2}}{(4\pi/3) r_{-2}^3}
            &=
            C\rho_\mathrm{crit}(z_{-2}),
        \end{align}
        where $C\simeq 650$ and $\rho_\mathrm{crit}(z_{-2})$ is the critical density at the redshift when the total mass in halo progenitors more massive than $f=0.02$ times the final halo mass was $M_{-2}$. According to EPS theory, this means
        \begin{align}\label{ludlow2}
            \frac{M_{-2}}{M_0}
            &=
            \erfc\!\left[\frac
            {\dc/D(z_{-2})-\dc/D(z_0)}
            {\sqrt{2[\sigma^2(fM_0)-\sigma^2(M_0)]}}
            \right],
        \end{align}
        where $M_0$ is the halo virial mass at the redshift $z_0$.
        With the Einasto profile of equation~(\ref{einasto}) specified, equations (\ref{ludlow1}) and~(\ref{ludlow2}) can be solved simultaneously to yield $r_{-2}$ and hence the halo concentration parameter $c$.
        Figure~\ref{fig:c} shows halo concentration parameters evaluated using this model for a range of masses and redshifts (dotted curves).
        Here we adopt \citet{2020A&A...641A...6P} cosmological parameters and use the \citet{1998ApJ...496..605E} fitting function to evaluate the matter power spectrum.
        
        For future subhalos, we can simply substitute into equation~(\ref{ludlow2}) the bias model
        \begin{align}
            \dc\to[1-D(z)/D(z_\mathrm{infall})]^{1.20+0.14a}\dc
        \end{align}
        determined in section~\ref{sec:modelLCDM} for the $M_{200\mathrm{c}}$ mass definition.
        We take $z_\mathrm{infall}=z_0$, so that we consider subhalos immediately prior to infall. The solid curves in figure~\ref{fig:c} show the concentration parameters of near-future subhalos evaluated in this way.
        Consistent with the findings of \citet{2025arXiv251204156D}, near-future subhalos have about 10--15\% higher concentration parameters than average field halos of the same mass at the same time.

		\section{Conclusion}\label{sec:conclusion}

        Halos destined to fall into a more massive host have biased growth histories before infall. Their main-progenitor masses and full progenitor mass functions both lie above those of typical field halos of the same mass at the same time.
        
        We have characterized this pre-infall bias compactly through a function $\beta(D/D_\mathrm{infall},a)$, where $D$ is the linear growth factor at scale factor $a$ and $D_\mathrm{infall}$ is the growth factor at infall. Within EPS theory, the bias corresponds to a modified collapse threshold $\dc\to\beta\dc$ for future subhalos. Calibrated against scale-free simulations and a concordance cosmology, $\beta$ takes the simple form $\beta(x,a)=(1-x)^{p+qa}$, with $p=1.20$ from the scale-free fit. The $qa$ term accounts for the late-time influence of dark energy; we find $q=0.14$ for the $M_{200\mathrm{c}}$ mass definition and $q=0.05$ for the $M_{200\mathrm{m}}$ mass definition.
        
        The bias propagates into subhalo properties. Because halos that assembled earlier are more centrally concentrated, future subhalos are more concentrated than typical field halos of the same mass at the same time. Applying our model to the concentration-mass relation of \citet{2016MNRAS.460.1214L} yields a 10--15\% concentration excess shortly before infall, consistent with the direct measurements of \citet{2025arXiv251204156D}. More generally, the assembly history bias should be incorporated into subhalo population models, which currently apply no pre-infall correction.

        The pre-infall subhalo bias is the extreme case of the more general phenomenon of assembly bias. In the EPS picture, $x = D/D_\mathrm{infall}$ has a direct interpretation: it is the normalized linear-theory overdensity $\delta_\mathrm{env}/\dc$ of the surrounding region at progenitor time $D$, since $\delta_\mathrm{env}=\dc D/D_\mathrm{infall}$ for a region collapsing at $D_\mathrm{infall}$. Recast with $x = \delta_\mathrm{env}/\dc$, the effective $(1-x)^{p+qa}\dc$ threshold might apply in general to halos in dense environments, not only to future subhalos. We leave further exploration of this point to future work.

		\appendix

        \section{No subhalo bias from excursion set theory with top-hat window}\label{sec:excursionset}
        
        Here we show that excursion set theory cannot predict the subhalo bias, even when correlations between steps are correctly accounted for. We use explicit calculations similar to those of \citet{2024MNRAS.528.1372D}. To set up the problem, suppose we are interested in a halo of mass $M_\mathrm{s}$ merging onto a larger halo of mass $M_\mathrm{h}$ at linear growth factor $D_\mathrm{infall}$. In the language of excursion set theory, this corresponds to a trajectory $\delta(M)$ that is constrained such that $\delta(M_\mathrm{s})=\delta(M_\mathrm{h})=\dc/D_\mathrm{infall}$, $\delta(M)<\dc/D_\mathrm{infall}$ for $M_\mathrm{s}<M<M_\mathrm{h}$, and $\delta(M)<\dc/D_\mathrm{infall}$ for $M>M_\mathrm{h}$.
        Note that these constraints imply $\left.\diff\delta(M)/\diff M\right|_{M=M_\mathrm{h}}=0$.
		
		The constraints $\delta(M_\mathrm{s})=\delta(M_\mathrm{h})=\dc/D_\mathrm{infall}$ and $\left.\diff\delta(M)/\diff M\right|_{M=M_\mathrm{h}}=0$ can be implemented straightforwardly in the calculation of an excursion set trajectory.
		Let $R$, $R_\mathrm{h}$, and $R_\mathrm{s}$ be the Lagrangian radii enclosing masses $M$, $M_\mathrm{h}$, and $M_\mathrm{s}$, respectively. Now discretize $R$ into $0<R_1<R_2<...<R_N$ for some finite $N$. The trajectory is now a vector $(\delta_{R_1},...,\delta_{R_N})$. Given the constraints, it has mean 
        \begin{align}
			\langle\delta_{R_i}|\delta_{R_\mathrm{s}},\delta_{R_\mathrm{h}},\frac{\diff\delta_{R_\mathrm{h}}}{\diff R_\mathrm{h}}\rangle
			&=
			\begin{pmatrix}
				\langle \delta_{R_i}\delta_{R_\mathrm{s}}\rangle &
				\langle \delta_{R_i}\delta_{R_\mathrm{h}}\rangle &
				\langle \delta_{R_i}\frac{\diff\delta_{R_\mathrm{h}}}{\diff R_\mathrm{h}}\rangle
			\end{pmatrix}
			\begin{pmatrix}
				\langle \delta_{R_\mathrm{s}}^2\rangle & \langle \delta_{R_\mathrm{s}}\delta_{R_\mathrm{h}}\rangle & \langle \delta_{R_\mathrm{s}}\frac{\diff\delta_{R_\mathrm{h}}}{\diff R_\mathrm{h}}\rangle \\ 
				\langle \delta_{R_\mathrm{s}}\delta_{R_\mathrm{h}}\rangle & \langle \delta_{R_\mathrm{h}}^2\rangle & \langle \delta_{R_\mathrm{h}}\frac{\diff\delta_{R_\mathrm{h}}}{\diff R_\mathrm{h}}\rangle \\
				\langle \delta_{R_\mathrm{s}}\frac{\diff\delta_{R_\mathrm{h}}}{\diff R_\mathrm{h}}\rangle & \langle \delta_{R_\mathrm{h}}\frac{\diff\delta_{R_\mathrm{h}}}{\diff R_\mathrm{h}}\rangle & \langle (\frac{\diff\delta_{R_\mathrm{h}}}{\diff R_\mathrm{h}})^2\rangle
			\end{pmatrix}^{-1}\!
			\begin{pmatrix}
				\dc/D_\mathrm{infall} \\ 
				\dc/D_\mathrm{infall} \\
				0
			\end{pmatrix}
		\end{align}
		and covariance matrix
		\begin{align}
			\AtBeginEnvironment{pmatrix}{\setlength{\arraycolsep}{1.0pt}}
			&\langle\Delta\delta_{R_i}\Delta\delta_{R_j}|\delta_{R_\mathrm{s}},\!\delta_{R_\mathrm{h}},\!\frac{\diff\delta_{R_\mathrm{h}}}{\diff R_\mathrm{h}}\rangle
			=
			\langle\delta_{R_i}\delta_{R_j}\rangle
			\!-\!
			\begin{pmatrix}
				\langle \delta_{R_i}\delta_{R_\mathrm{s}}\rangle &
				\langle \delta_{R_i}\delta_{R_\mathrm{h}}\rangle &
				\langle \delta_{R_i}\frac{\diff\delta_{R_\mathrm{h}}}{\diff R_\mathrm{h}}\rangle
			\end{pmatrix}\!\!
			\begin{pmatrix}
				\langle \delta_{R_\mathrm{s}}^2\rangle & \langle \delta_{R_\mathrm{s}}\delta_{R_\mathrm{h}}\rangle & \langle \delta_{R_\mathrm{s}}\frac{\diff\delta_{R_\mathrm{h}}}{\diff R_\mathrm{h}}\rangle \\ 
				\langle \delta_{R_\mathrm{s}}\delta_{R_\mathrm{h}}\rangle & \langle \delta_{R_\mathrm{h}}^2\rangle & \langle \delta_{R_\mathrm{h}}\frac{\diff\delta_{R_\mathrm{h}}}{\diff R_\mathrm{h}}\rangle \\
				\langle \delta_{R_\mathrm{s}}\frac{\diff\delta_{R_\mathrm{h}}}{\diff R_\mathrm{h}}\rangle & \langle \delta_{R_\mathrm{h}}\frac{\diff\delta_{R_\mathrm{h}}}{\diff R_\mathrm{h}}\rangle & \langle (\frac{\diff\delta_{R_\mathrm{h}}}{\diff R_\mathrm{h}})^2\rangle
			\end{pmatrix}^{\!\!\!\!-1}\!\!\!\!
			\begin{pmatrix}
				\langle \delta_{R_j}\delta_{R_\mathrm{s}}\rangle \\ 
				\langle \delta_{R_j}\delta_{R_\mathrm{h}}\rangle \\
				\langle \delta_{R_j}\frac{\diff\delta_{R_\mathrm{h}}}{\diff R_\mathrm{h}}\rangle
			\end{pmatrix}\!\!.
			\AtBeginEnvironment{pmatrix}{\setlength{\arraycolsep}{2.0pt}}
		\end{align}
		The relevant expectation values can be evaluated as integrals over the matter power spectrum by noting that the average density contrast $\delta_R(\vec x)$ in a sphere of radius $R$ centered on position $\vec x$ satisfies
        \begin{align}\label{deltaR}
        	\delta_R(\vec x)&=\int\frac{\diff^3\vec k}{(2\pi)^3}\e^{\I \vec k\cdot\vec x}\delta(\vec k)W(kR)
        	\qquad
        	\text{and}
        	\qquad
        	\frac{\diff}{\diff R}\delta_R(\vec x)=\int\frac{\diff^3\vec k}{(2\pi)^3}\e^{\I \vec k\cdot\vec x}\delta(\vec k)k W'(kR)
        \end{align}
        where $W(x)\equiv (3/x^3)(\sin x-x\cos x)$ is the spherical top-hat window function in Fourier space. See \citet{2024MNRAS.528.1372D} for details.
        
        We adopt a \citet{1998ApJ...496..605E} matter power spectrum \citep[using cosmological parameters from ][]{2020A&A...641A...6P}, discretize $M$ into $10^3$ values logarithmically spaced between $1~\Msol$ and $10^{15}~\Msol$, and consider several combinations of the subhalo mass $M_\mathrm{sub}=M_\mathrm{s}$, host mass $M_\mathrm{host}=M_\mathrm{h}$, and infall redshift $z_\mathrm{infall}$. We randomly sample trajectories $\delta(M)$ using the methods of \citet{2024MNRAS.528.1372D} and discard any trajectories that fail to satisfy $\delta(M)<\dc/D_\mathrm{infall}$ for all $M_\mathrm{s}<M<M_\mathrm{h}$ and all $M>M_\mathrm{h}$. We adopt $\dc=1.5$ following \citet{2024MNRAS.528.1372D}, but this has no impact on our results. We continue until we have sampled $3\times 10^4$ trajectories for each scenario.
        
\begin{figure*}
			\centering
			\includegraphics[width=\linewidth]{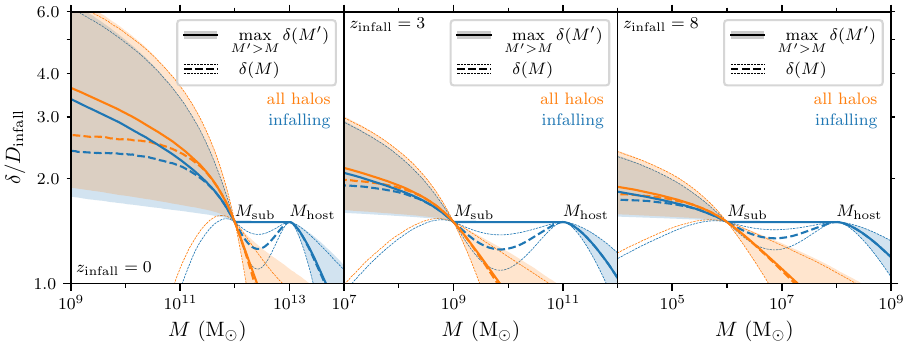}
			\caption{
            The blue curves represent trajectories in excursion set theory for mass elements that transition from residing in a halo of mass $M_\mathrm{sub}$ to residing in a halo of mass $M_\mathrm{host}\gg M_\mathrm{sub}$ at redshift $z_\mathrm{infall}$.
            For comparison, the orange curves show trajectories without the $z_\mathrm{infall}$ conditioning.
            The thick and thin dashed curves show the median and 16th/84th percentiles of $\delta(M)$, respectively, while the solid curves and shading show the median and 16th/84th percentiles of the downward-running maximum of $\delta(M)$, which is the more relevant quantity for halo assembly histories in excursion set theory.
            }
			\label{fig:trajectories}
		\end{figure*}

        Figure~\ref{fig:trajectories} shows the $\delta(M)$ trajectories sampled in this manner (blue). We show both the median and 16th and 84th percentiles of $\delta(M)$ (dashed curves) along with the same quantiles of $\max_{M'>M}\delta(M')$ (solid curves and shading). The latter are more relevant for halo growth in the context of excursion set theory. Specifically, each trajectory corresponds to a dark matter particle, and at the linear growth factor $D$, the mass $M$ of the halo hosting the particle satisfies
        \begin{align}
        	\dc/D=\max_{M'>M}\delta(M').
        \end{align}
        At fixed mass $M$, trajectories with higher values of $\max_{M'>M}\delta(M')$ correspond to particles whose halos achieved the mass $M$ earlier.
        
		For comparison, we also sample $3\times 10^4$ trajectories with the relaxed constraints $\delta(M_\mathrm{s})=\dc/D_\mathrm{infall}$ and $\delta(M)<\dc/D_\mathrm{infall}$ for all $M>M_\mathrm{s}$. These are the general field-halo analogues -- they are constrained to produce a halo of mass $M_\mathrm{s}$ at the same time $D_\mathrm{infall}$ but are not further constrained to infall into a halo of mass $M_\mathrm{h}$ at that time. These trajectories are also shown in figure~\ref{fig:trajectories} (orange).
		
		In figure~\ref{fig:trajectories}, the unconstrained trajectories (orange) tend to lie above the constrained trajectories (blue) at any fixed mass $M<M_\mathrm{s}$. Within the excursion set theory interpretation, this would imply that general field halos tend to have slower growth histories, with more mass accreted earlier, relative to future subhalos. This is exactly the opposite trend from the subhalo bias seen in simulations.
		Hence, excursion set theory with correlated steps (assuming a top-hat window $W$) cannot explain the subhalo bias.
        
		\bibliography{main}{}
		\bibliographystyle{aasjournalv7}
		
		
		
	\end{document}